\documentclass[a4paper,fleqn,usenatbib]{mnras}

\usepackage{newtxtext,newtxmath}

\usepackage[T1]{fontenc}
\usepackage{ae,aecompl}

\usepackage{graphicx}	
\usepackage{amsmath}	
\usepackage{amssymb}	

\def\spose#1{\hbox to 0pt{#1\hss}}
\def\lta{\mathrel{\spose{\lower 3pt\hbox{$\mathchar"218$}}
     \raise 2.0pt\hbox{$\mathchar"13C$}}}
\def\gta{\mathrel{\spose{\lower 3pt\hbox{$\mathchar"218$}}
     \raise 2.0pt\hbox{$\mathchar"13E$}}}

\def\p0{\phantom{0}}

\title[Comets in 55 Cancri]
{Theoretical Studies of Comets in the 55 Cancri System}

\author[]{
Rudolf Dvorak$^{1}$, Birgit Loibnegger$^1$ and Manfred Cuntz$^{2}$\thanks{E-mail: cuntz@uta.edu (MC)}
\\
$^{1}$Institute of Astrophysics, University of Vienna, T\"urkenschanzstra{\ss}e 17, A-1180 Vienna, Austria \\
$^{2}$Department of Physics, University of Texas at Arlington, Arlington, TX 76019, USA
}

\date{Accepted XXX. Received YYY; in original form ZZZ}

\pubyear{2020}

\begin{document}
\label{firstpage}
\pagerange{\pageref{firstpage}--\pageref{lastpage}}
\maketitle

\begin{abstract}
We present orbital integrations for various
Jupiter family comets (JFCs) in the 55~Cancri system.  This star is known to possess (at least)
five planets with masses ranging from super-Earth to Jupiter-type.  Furthermore, according to
observational constraints, there is a space without planets between $\sim$0.8~au and $\sim$5.7~au,
offering the principal possibility of habitable terrestrial planets, including long-term orbital stability.
Hence, there is a stark motivation for the study of comets in the 55~Cnc system noting that comets
are viewed a viable candidate mechanism for the delivery of water to Earth-type planets
located in stellar habitable zones.  However, our study shows that the duration of stability of
JFC analogs in the 55~Cnc system is much shorter compared to comets in the Solar System owing to
profound differences in the planetary structure of the systems.  For example, between planet
55~Cnc-f and 55~Cnc-d, the comets do not undergo close cometary encounters akin to Earth and Mars in the
Solar System as the planetary masses in the 55~Cnc system are much larger than in the Solar System
and therefore the comets are much less orbitally stable.  Nevertheless, we expect an increased number
of comet--planet encounters as well as cometary collisions in the 0.8 / 5.7~au gap.  Future
observations and additional theoretical studies are required to shed light on the possibility of
habitable terrestrial planets in the 55~Cnc system, including the possible role(s) of exocomets
in the facilitation of planetary habitability.
 \end{abstract}

\begin{keywords}
astrobiology -- circumstellar matter -- comets: general --
methods: numerical -- protoplanetary disks -- stars: individual (55~Cnc)
\end{keywords}



\section{Introduction}\label{sec:intro}

An active field of research is the study of comets around stars other than the Sun, also referred to as exocomets.
Previous observational results have been given by, e.g., \cite{beu90}, \cite{kie14}, \cite{eir16}, and \cite{rap18}.
Important topics of interest relevant to the fields of astrophysics and astrobiology include (1) studies on the
origin of comets in different types of systems, (2) gravitational interaction of comets with respect to stars and
their planetary systems and (3) the proliferation of water and possible biomolecules to Earth-type planets, including
planets situated in the stellar habitable zones; see, e.g., \cite{ste87} and \cite{ray17} for background information.
Recently, the research group centered at Vienna, Austria, made various contributions to that field, including model
simulations for the systems of HD~10180, Proxima Centauri, 47~UMa, and HD~141399; see \cite{loi17}, \cite{sch18},
\cite{cun18}, and \cite{dvo20}, respectively.

Here we investigate the extrasolar planetary system 55~Cnc with a particular focus on (analogs of) Jupiter family
comets (JFC) found in our Solar System.  Previous studies on the origin and dynamics of Solar System comets
have been given by, e.g., \cite{lev97}, \cite{eme13}, and \cite{fou17,fou18}.
An important motivation for our efforts stems from the fundamentally
different structures of the two systems as identified by the different masses and star--planet distances of
the system planets.  Although 55 Cancri is a stellar binary with its components
readily referred to as 55~Cnc~A and 55~Cnc~B, we do not consider the companion 55~Cnc~B
in our integrations owing to its small mass ($\sim$0.25 $M_{\odot}$;
see \citealt{gai14} and \citealt{new17}) and the large separation distance of the two stellar components
given as about 1000~au \citep{duq91}.  In fact, the influence of 55~Cnc~B (and of any planet hosted by that
component) on the cometary orbits associated with the star-planet system of 55~Cnc~A is negligible.

Further simplifications are discussed below in more detail.  Specifically, in our numerical simulations
we only include the planets 55~Cnc-c, 55~Cnc-f, and 55~Cnc-d (see Table~\ref{tab:planets_param}).
The other two planets, i.e., 55~Cnc-b and 55~Cnc-e, orbit the star in relatively close proximity
($d < 0.12$~au), which makes cometary encounters with these planets highly improbable.
This view receives support from checking the respective perihelion distances in the list of all JFCs
(i.e., $\omega < 0.336$ au except for three
SOHO\footnote{The SOlar and Heliospheric Observatory (SOHO) studies the internal structure of
the Sun as well as its extended outer atmosphere including the origin of the solar wind.}
comets, which are 321P, 322P, and 323P with $\omega \approx 0.05$~au, with $\omega$ being the pericentric distance).
Moreover, most of the JFCs have their closest encounters to the Sun outside Earth's orbit.

In this work, we focus on 55~Cancri (55~Cnc, $\rho^1$~Cnc), a G8~V star \citep{gon98}, with
an effective temperature\footnote{Further determinations of stellar parameters for 55~Cnc have
been given by \cite{bra11}; they are consistent with those adopted in this study.}
and luminosity lower than those of the Sun \citep[e.g.,][]{fis05,lig16}.  Note that
55~Cnc is considerably older than the Sun with an estimated age of 8~Gyr \citep{mam08}.
Moreover, 55~Cnc has a mass of 0.96~$M_\odot$ \citep{lig16}; it is therefore considered an orange dwarf.
Additionally, 55~Cnc is known to host five planets, discovered between 2008 and 2011 by \cite{fis08},
\cite{daw08}, and \cite{win11} based on the radial velocity method.  All of these planets have masses
significantly larger than Earth.  However, based on previous theoretical studies, see, e.g.,
\cite{blo03}, \cite{riv07}, \cite{ray08}, \cite{smi09}, and \cite{sat19}, the existence of Earth-mass
planets in that system is highly plausible.

Studies of 55~Cnc, as well as of similar stars, are of particular interest to astrobiology in consideration of
the various favorable properties of general orange dwarfs, i.e., late-type G and early-type K main-sequence stars
\citep[e.g.,][]{cun16,lin18,lin19,sch19}.  These stars are considered particularly suitable for hosting planets
with exolife in light of the numerous features supportive of exobiology, including (but not limited to)
the frequency of those stars, the relatively large size of their HZs (if compared to M dwarfs), and
their relatively long main-sequence life times (i.e., 15~Gyr to 30~Gyr, compared to $\sim$10~Gyr for
solar-like stars).  A detailed analysis of the physical constraints on the likelihood of life on exoplanets
in various stellar environments has been given by \cite{lin18}.  Note that 55~Cnc is expected to possess many of
those features.

Our paper is structured as follows: In Section~2, we present our methods and the numerical setup.
In Section~3, we convey our results and discussion, including studies of the (hypothetical) `Jupiter family'
in 55~Cnc as well as simulations for comet analogs of the Solar System comets 2P~/~Encke, 9P~/~Tempel~1, and
31P~/~Schwassmann-Wachmann~2.  Our conclusions and comments regarding possible future research are given in
Section~4.


\section{Methods and Numerical Setup}

For our study we assume that 55~Cancri harbors an Oort-type cloud of comets
at the system's outskirts.  These comets are expected to be gravitationally disturbed by,
e.g., passing stars or through galactic tidal forces, resulting in highly eccentric cometary
orbits, thus allowing the comets to enter the system's inner domain.  Cometary trajectories,
including their long-term developments, as well as mechanisms of cometary injections into the
system have previously been studied by numerous authors \cite[e.g.,][and references
therein]{fou17,ric17,dvo20}.

The goal of the current study is to explore the dynamics of comets,
including their interaction with the planets of the 55~Cancri system.
We will examine the dynamics of some Solar System comet analogs\footnote{Solar comets put
into the 55~Cnc system are referred to as {\it comet analogs}; their initial values, except
for their semimajor axes, have been chosen to match those of the respective solar system comet.
{\it Comet clones} denote sets of comets created through variations of their $MA$ values,
which is done for the purpose of detailed numerical simulations.}, namely, 2P~/~Encke,
9P~/~Tempel~1, and 31P~/~Schwassmann-Wachmann~2; they are later referred to as comet A, B,
and C, respectively, for convenience.  We will also include a fictitious Earth-mass planet
in order to investigate its influence on the trajectories of the comets.
This planet is put in a stable orbit with a semimajor axis of 1~au.

For information on the 55~Cancri system including its five planets, see Table~\ref{tab:planets_param};
However, contrary to the recent study by \cite{sat19}, we include the two inner planets in a limited sense
only by adding their masses to the mass of the central star.  This is a common practice routinely
also adopted for the integration of the Solar System.  In this case, Mercury's mass is added to the
Sun's mass when the desired precision is not at the limit of the integration itself (1.0E-15 for
double precision computations).  However, regardless of the process, the various planetary parameters
for the 55~Cnc system, used as initial conditions, are not very well constrained, especially the
planetary masses are in fact minimum masses due to their derivation from radial velocities observations.

Our method is based on a direct numerical integration with Lie-series of the equations of motion, 
a method readily used for computations for $N$-body integrations --- it has already been compared
to other methods for numerical integration in various papers \citep[e.g.,][]{han84}.
This method solves the differential equations by differentiation that leads to a power series
with factors computed in advance.  This approach is quite fast and competitive compared to those used
by many other authors.  In addition, the length of the time-step can be changed separately for every step;
this approach increases the precision of the method, which is particularly helpful for close encounters
and collisions \citep[e.g.,][]{dvo20}.  Based on the data of Table~\ref{tab:planets_param}, the two
innermost planets 55~Cnc-e and 55~Cnc-b with orbital periods of less than a day and $\sim$15 days,
respectively, would entail very small step sizes for the integration.  Therefore, their masses
have been added to the mass of the central star.

%

\begin{table*}
	\centering
	\caption{Orbital Parameters of the 55 Cancri Planets}
\begin{tabular}{lccccccc}
\noalign{\smallskip}
\hline
\noalign{\smallskip}
Planet           & Mass         & $a$   & $e$ & $i$ & $\omega$ & $MA$ & Orbital Period \\
...              & ($M_\oplus$) & (au)  & ... & ($^\circ$) &($^\circ$) & ($^\circ$) & (days)   \\
\noalign{\smallskip}
\hline
\noalign{\smallskip}
55 Cancri e & 8.37   & 0.0156 & 0.170 & 7.5 &  90 & 204.032 & 0.7365 \\
55 Cancri b & 264.75 & 0.1148 & 0.010 & 0.0 & 110 &  19.88  & 14.650 \\
55 Cancri c & 54.38  & 0.2403 & 0.005 & 0.0 & 356 & 116.10  & 44.364 \\
55 Cancri f & 57.209 & 0.781  & 0.320 & 0.0 & 139 & 198.61  & 259.80 \\
55 Cancri d & 1169.6 & 5.74   & 0.020 & 0.0 & 254 & 10.89   & 5169.0 \\
``Earth''   & 1.0    & 1.0    & 0.0   & 0.0 & 0.0 &  0.0    & ...    \\
\noalign{\smallskip}
\hline
\noalign{\smallskip}
\multicolumn{8}{p{1.15\columnwidth}}{
Note:  The planetary parameters (mass, $a$, $e$, $i$, $\omega$, and $MA$, with all parameters having their
usual meaning) of the 55 Cancri planets are listed, see references \cite{fis08}, \cite{daw08}, \cite{win11},
\cite{end12}, and \cite{lig16}).  The ascending node $\Omega$ is set to zero for all planets.
}
\end{tabular}
\label{tab:planets_param}
\end{table*}


\section{Results and Discussion}

\subsection{The `Jupiter family' in 55 Cancri}

The 580 JFC examined here have been placed into the 55~Cnc system with their orbital elements
adopted from the JPL Small-Body Database\footnote{For information see link:
{\tt ssd.jpl.nasa.gov/sbdb.cgi}.}  The focus of our work is to study various Solar System
comet system analogs in further detail.
However, the values for their semimajor axes have been normalized based on Jupiter's
semimajor axis and that of 55~Cnc-d.  The other elements, i.e., the
cometary eccentricities and inclinations (see Table~\ref{tab:orb_info}) were left unchanged.
For each comet analog we created 100 initial conditions based on slight variations
in their initial mean anomalies $MA$, bracketed by $MA=0^{\circ}$ and $MA=360^{\circ}$ with
a step size of $\delta MA=3.6^{\circ}$, which led to a total of almost 60,000 integrated orbits.
The total integration time for each orbit was set to 100~kyr.

\begin{figure*}
\centering
  \includegraphics[width=0.65\linewidth]{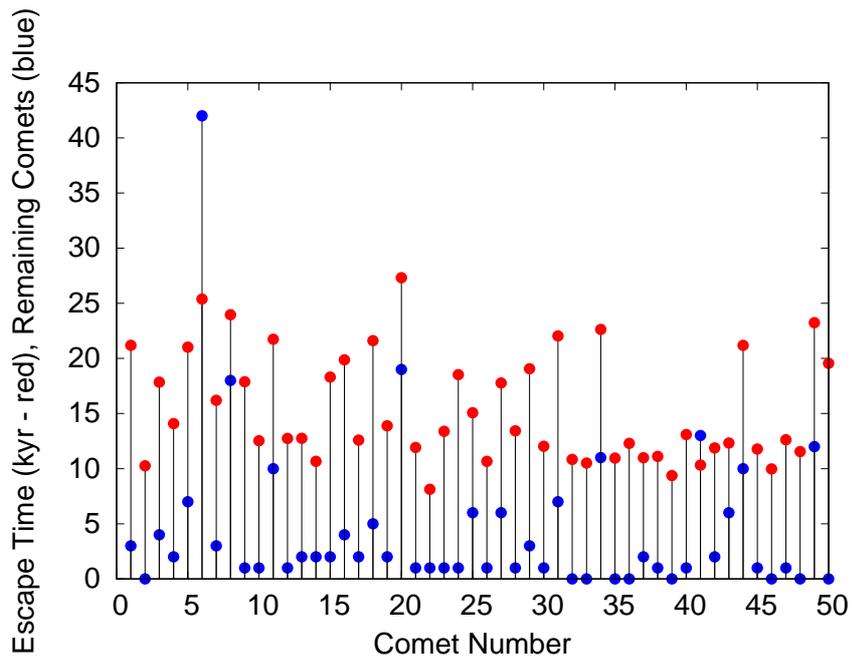}
  \caption{Information on comets no. 1 -- 50.  Special cases: 1 comet~A, 6 comet~B, 22 comet~C}
  \label{fig:100}
\end{figure*}

\begin{figure*}
  \includegraphics[width=0.49\linewidth]{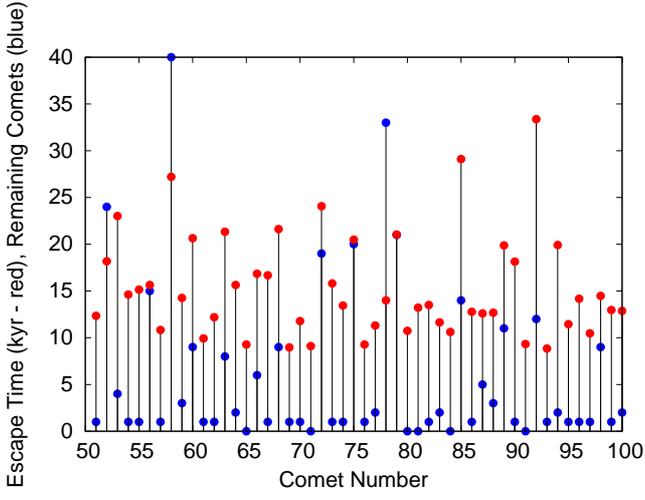}
  \includegraphics[width=0.49\linewidth]{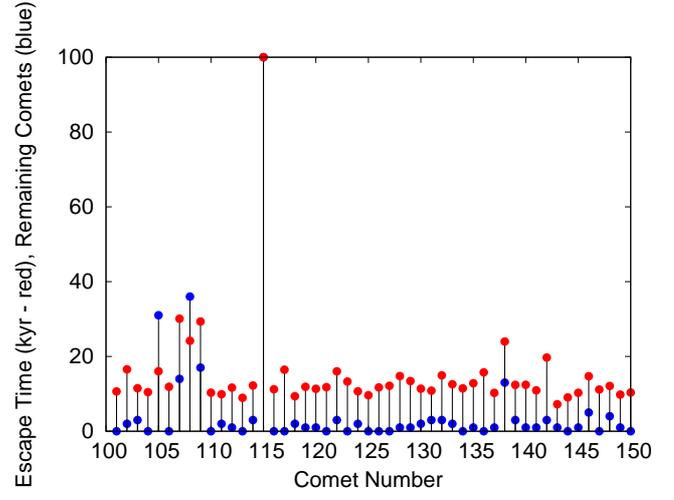}
  \caption{Information on comets no. 51 -- 100 and no. 101 -- 150.}
  \label{fig:50-150}
\end{figure*}

\begin{figure*}
  \includegraphics[width=0.49\linewidth]{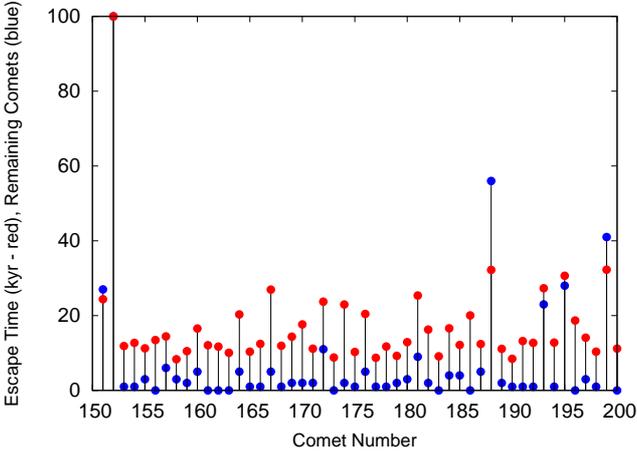} 
  \includegraphics[width=0.49\linewidth]{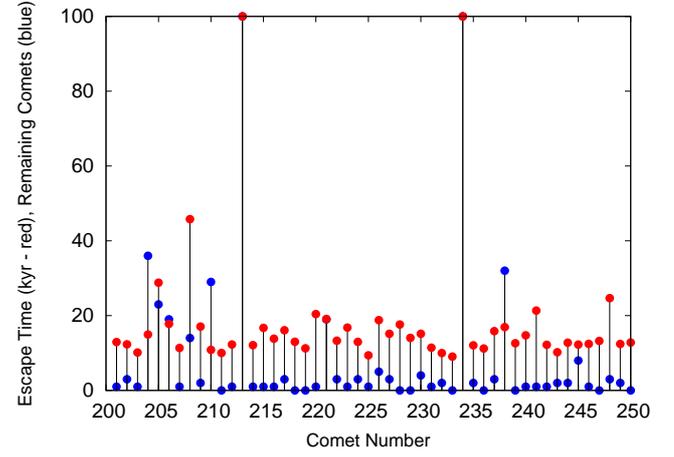}
  \caption{Information on comets no. 151 -- 200 and no. 201 -- 250.}
  \label{fig:150-250}
\end{figure*}

\begin{figure*}
  \includegraphics[width=0.49\linewidth]{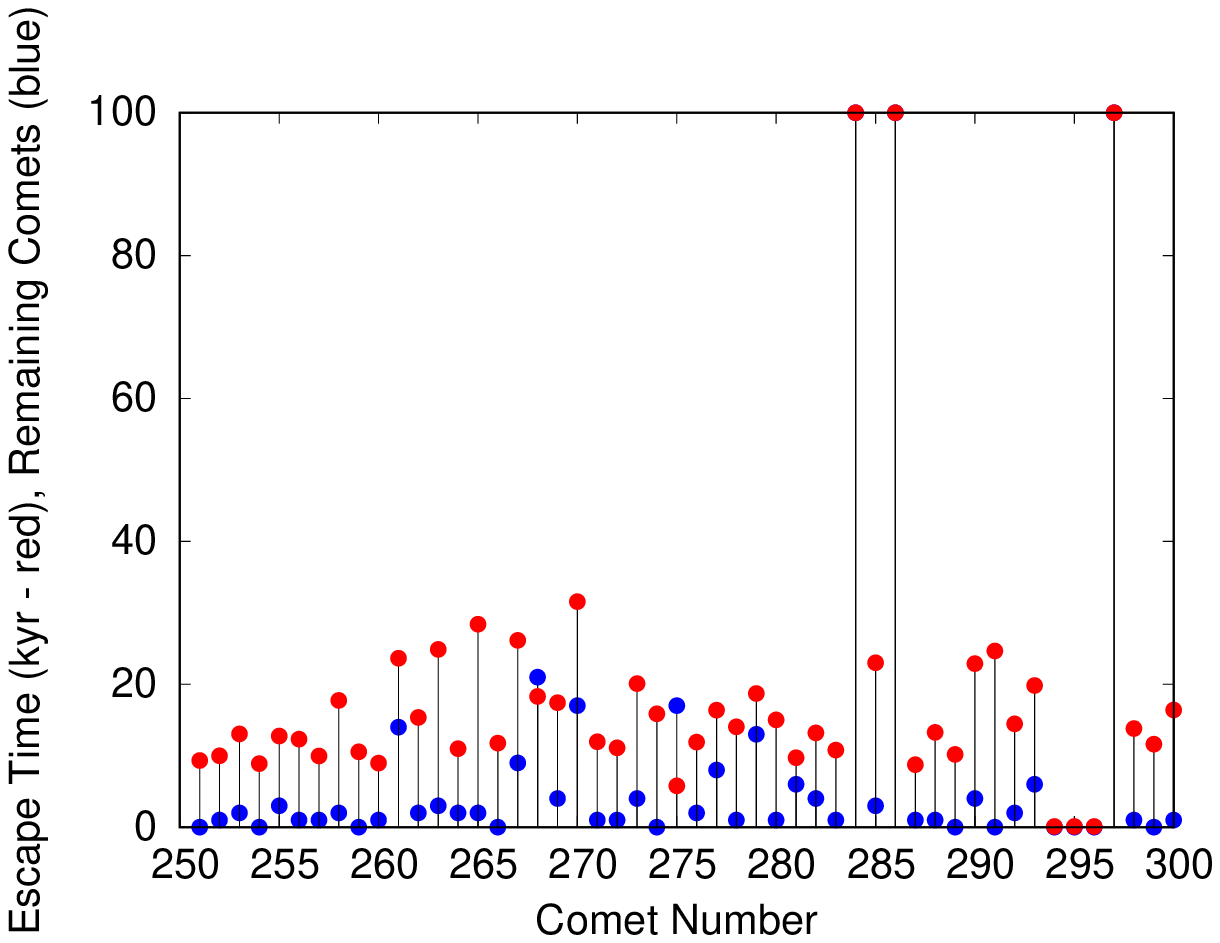} 
  \includegraphics[width=0.49\linewidth]{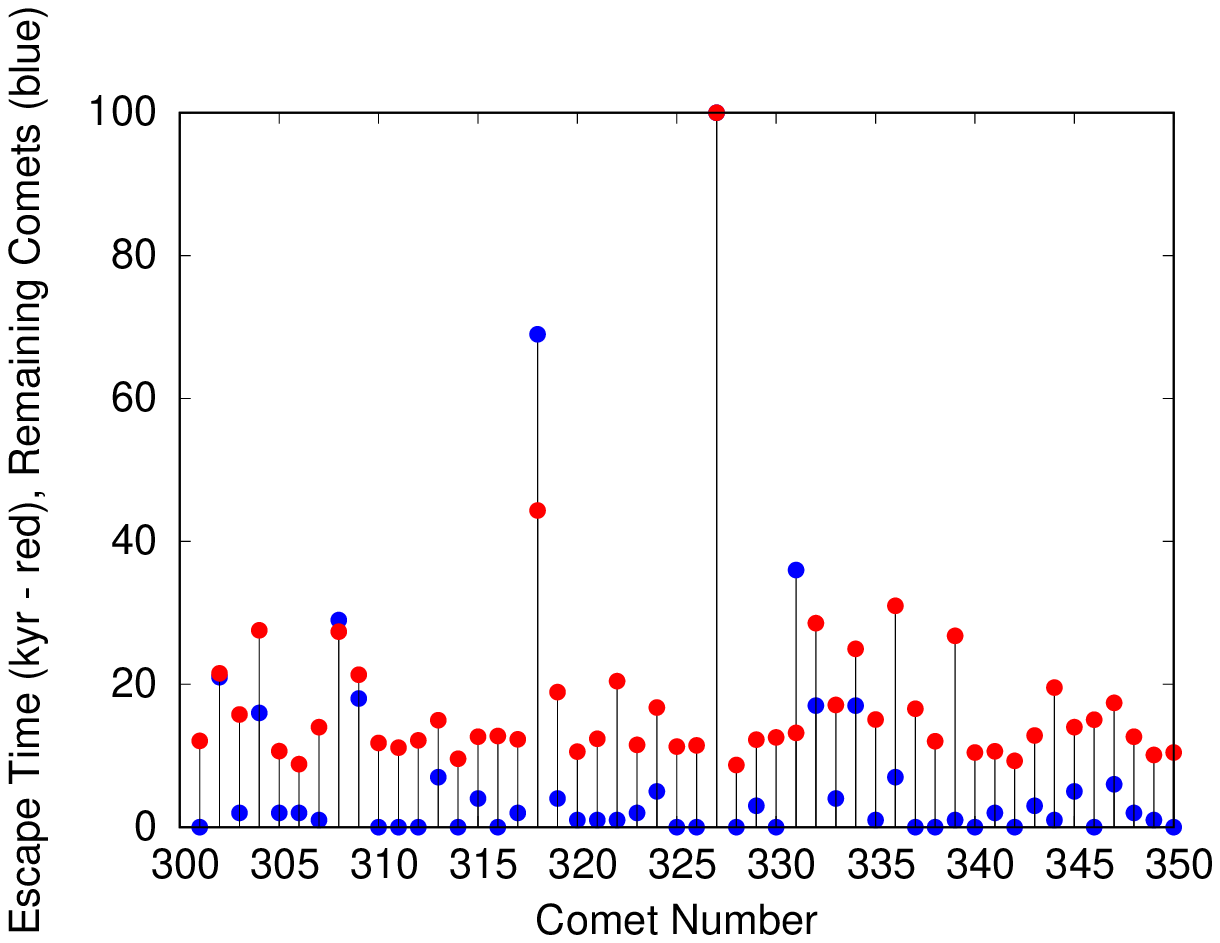}
  \caption{Information on comets no. 251 -- 300 and no. 301 -- 350.}
  \label{fig:250-350}
\end{figure*}

The results for each comet were analyzed with respect to their mean escape times and the
number of remaining comets, see Figures~\ref{fig:100} to \ref{fig:250-350}.  In the following,
we explain the plots while focusing on the results for three examples chosen in regard to a
large, medium, and small cometary eccentricity.  In order to further compare the dynamical
evolution for the three comets placed in the 55~Cnc system relative to the
Solar System we adopted two models: one that includes the planets Venus to
Saturn (Ve2Sa) and another one excluding Saturn (Ve2Ju); see Table~\ref{tab:escapers}.

%

\begin{table*}
	\centering
	\caption{Orbital Information for Solar System Comet Analogs}
\begin{tabular}{llcccc}
\noalign{\smallskip}
\hline
\noalign{\smallskip}
Name & Solar System Name  & Semimajor Axis  & Eccentricity & Inclination  & Tisserand Invariant \\
...  & ...                & (au)            & ...          & ($^{\circ}$) & ...                 \\
\noalign{\smallskip}
\hline
\noalign{\smallskip}
Comet~A  & 2P~/~Encke                   &  2.212 & 0.848 & 11.8  & 3.026  \\
Comet~B  & 9P~/~Tempel~1                &  3.138 & 0.514 & 10.5  & 2.969  \\
Comet~C  & 31P~/~Schwassmann-Wachmann~2 &  4.249 & 0.194 &  4.5  & 2.992  \\
\noalign{\smallskip}
\hline
\noalign{\smallskip}
\end{tabular}
\label{tab:orb_info}
\end{table*}

%

\begin{table*}
	\centering
	\caption{Outcome for the Solar System Clone Simulations}
\begin{tabular}{llllll}
\noalign{\smallskip}
\hline
\noalign{\smallskip}
Name    & Solar System Name               & Item & Ve2Sa & Ve2Ju & 55~Cnc \\
\noalign{\smallskip}
\hline
\noalign{\smallskip}
Comet~A  &  2P~/~Encke                    & $t_{\rm esc}$  & 82 kyr & 88 kyr & 21 kyr \\
Comet~A  &  2P~/~Encke                    & $N_{\rm esc}$  & 2      &  1     & 97     \\
Comet~B  &  9P~/~Tempel~1                 & $t_{\rm esc}$  & 55 kyr & 53 kyr & 25 kyr \\
Comet~B  &  9P~/~Tempel~1                 & $N_{\rm esc}$  & 42     & 48     & 58     \\
Comet~C  &  31P~/~Schwassmann-Wachmann~2  & $t_{\rm esc}$  & 40 kyr & 41 kyr & 8 kyr  \\
Comet~C  &  31P~/~Schwassmann-Wachmann~2  & $N_{\rm esc}$  & 43     & 54     & 98     \\
\noalign{\smallskip}
\hline
\noalign{\smallskip}
\multicolumn{6}{p{1.15\columnwidth}}{
Note:  Mean escape times $t_{\rm esc}$ and number of escapers $N_{\rm esc}$ out of 100 comet clones.
}
\end{tabular}
\label{tab:escapers}
\end{table*}

The outcome is that the lifetimes of the comets in 55~Cnc are significantly shorter compared to
the corresponding comets in the Solar System because of the large masses of 55~Cnc-f and 55~Cnc-d.
This is especially true for comet~A owing to the large eccentricities of the respective comet
clones; their close encounters to the 55~Cnc system planets expel most of them from the
system as only 3 out of 100 of the comet clones are able to survive.  This result is
drastically different for simulations of the Solar System where only a few of the
comet~A clones were ejected (with an ejection defined as $e$~$\ge$~1);
i.e., 2 and 1 for the Ve2Sa model and the Ve2Ju model, respectively.
After a total integration time of 100 kyr, only 42 of the 100 comet clones for 55~Cnc survived
for comet~B, and only 2 of the clones for comet~C.  The reason for the few remaining clones
for comets A and C is their close proximity to 55~Cnc-d, which in combination with the cometary
eccentricities result in close planetary encounters shortly after start of the integrations;
therefore, many of the comet clones are ejected soon after.  Hence, it is found that
the JFCs within the Solar System are much more stable compared to the corresponding comets
hosted by 55~Cnc.


\subsection{Lifetime of Comets}\label{life_time}

%

%

%

%

\subsubsection{General Aspects}\label{gen_aspects}

Figures~\ref{fig:100} - \ref{fig:250-350} depict the mean escape times (in red)
and the mean numbers of the remaining comet clones (out of 100; in blue) for various
case studies.  We omit showing the figures for the faint comets no.~351 to 580, which
are not significantly different from the previous ones.  Some comets do not escape during
the integration time of 100~kyr, which is attributable to the fact that these orbits
are stable for thousands of orbital periods.
Note that in some cases the red and blue filled circles overlap (e.g., comet 151
in the left panel of Figure~\ref{fig:150-250}).  In this particular case, the escape time
has been equal to the integration time for all 100 clones of this particular comet.
This means that all 100 clones are stable for the entire time of integration.

On the other hand,
in some cases there are no surviving comet clones in our simulations.  These kinds of orbits
are highly unstable due to their very large eccentricities of $e > 0.97$; in case of the Sun,
those types of comets have been discovered by SOHO.  In conclusion, we find that most of the
$\sim60,000$ comets have a mean escape time below 20~kyr and the mean number of surviving comets
remains in most cases below 40 of 100 clones. Table~\ref{tab:JFC_55Cnc} lists the stable JFCs
analogs for the 55~Cnc system based on simulations extending to 100~kyr.

%

\begin{table}
	\centering
	\caption{Stable JFC Analogs in 55~Cnc}
\begin{tabular}{l}
\noalign{\smallskip}
\hline
\noalign{\smallskip}
  176P~/~LINEAR               \\
  238P~/~Read                 \\
  259P~/~Garradd              \\
  311P~/~PanSTARRS            \\
  313P~/~Gibbs                \\
  324P~/~La Sagra             \\
  358P~/~PanSTARRS            \\
  P/2015~X6~~~(PanSTARRS)     \\
  P/2016~G1~~~(PanSTARRS)     \\
  P/2019~A4~~~(PanSTARRS)     \\
  P/2019~A7~~~(PanSTARRS)     \\
\noalign{\smallskip}
\hline
\noalign{\smallskip}
\end{tabular}
\label{tab:JFC_55Cnc}
\end{table}


\subsubsection{The Comet 2P~/~Encke}\label{examples1}

%

\begin{figure*}
  \includegraphics[width=0.35\linewidth, angle=270]{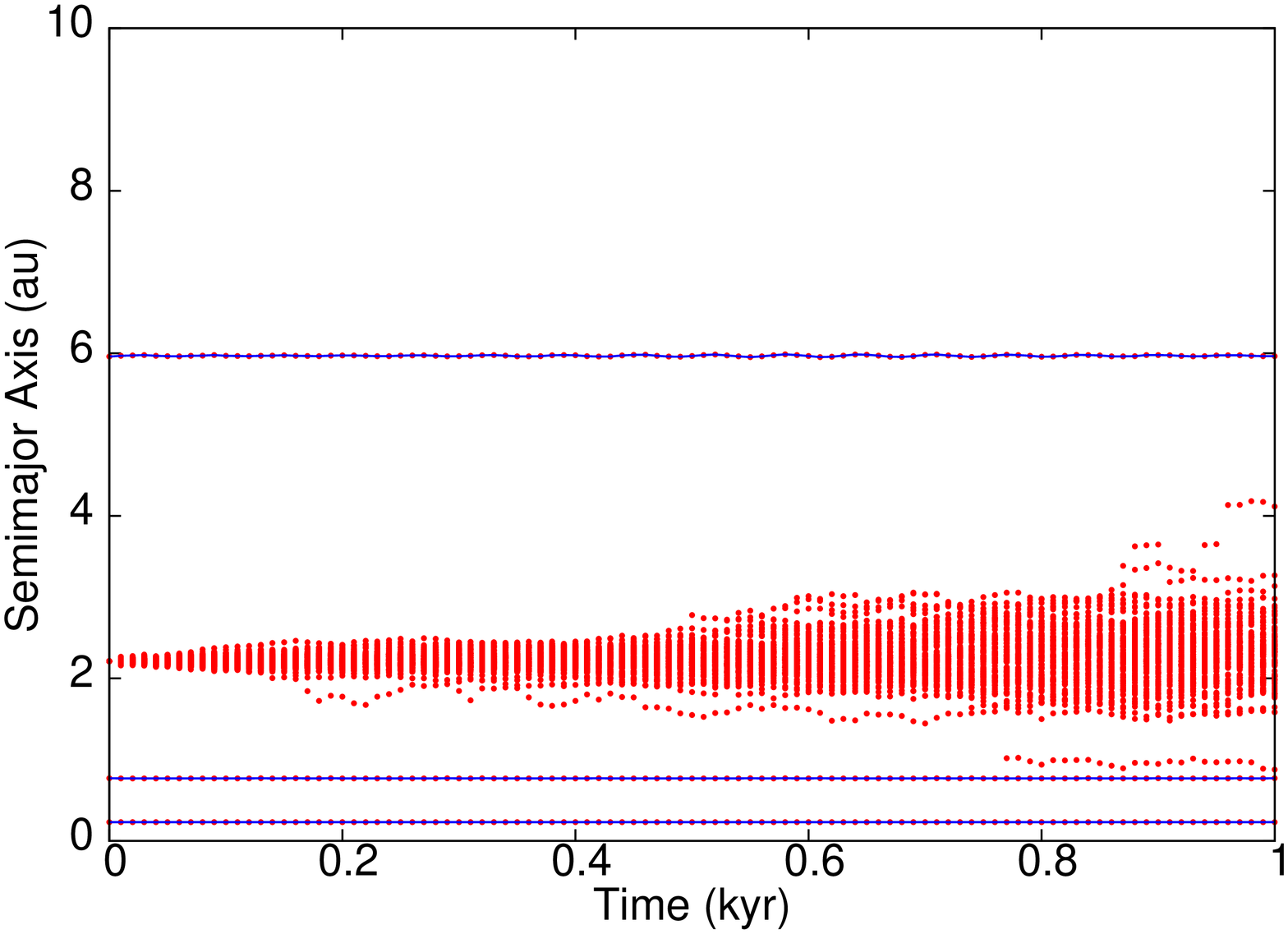}
  \includegraphics[width=0.35\linewidth, angle=270]{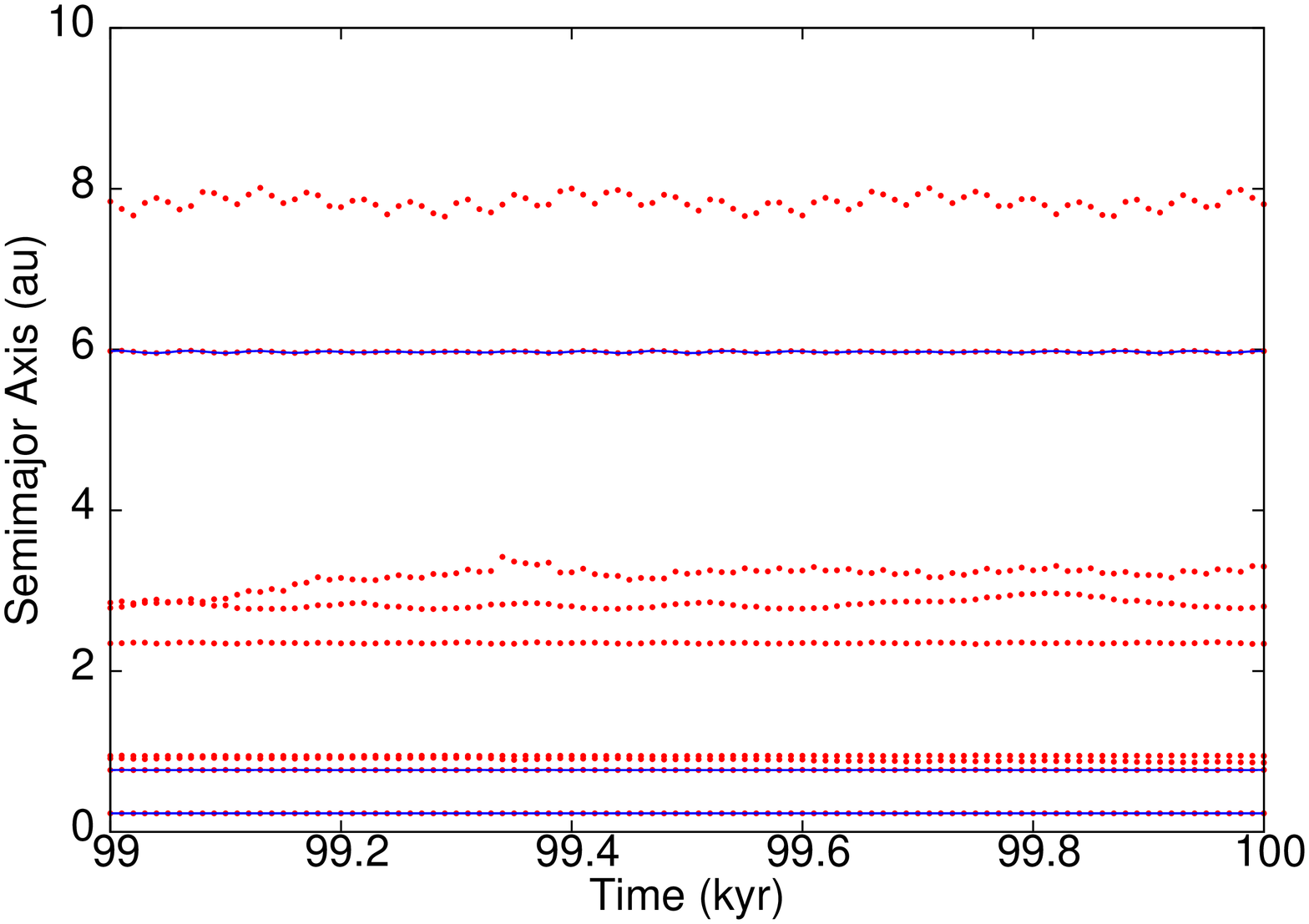}
  \caption{Temporal evolution of the clones of comet~A (2P~/~Encke).
    Left panel: evolution of the semimajor axes up to 1~kyr; right panel: evolution of the semimajor axes from 99~kyr to 100~kyr.
    Note that up to 1~kyr only a few clones remain captured close to the star.  The blue lines denote the planetary distances,
    whereas the red points represent the clones.}
  \label{fig:Encke}
\end{figure*}

Figure~\ref{fig:Encke} depicts the orbits of the clones for comet~A at the beginning and end of the integration.
One can see that after a short time, i.e., about 200~yr, the comet clones start experiencing the perturbations
by the planets; consequently, their semimajor axes increase leading to possible escapes from the system presumably
after close planetary encounters.  Only a few of the comet clones remain in the system toward the end of the integration.
One of the comets was pushed to an orbit beyond 55~Cnc-d at approximately 7.9~au where it remained stable
with a relatively high eccentricity.   In Table~\ref{table-Tempel} the orbital elements of the remaining clones for comet~A
are given.  One can see that remaining clone no.~4 almost retains its initial eccentricity; however, its inclination is
now quite large.  It was pushed to an outer orbit at the beginning of the integration and stayed there relatively unperturbed
until the integration was stopped.  Except for no.~6, all the other comet clones gained higher inclinations during the simulation
in compared to their initial values.

Comparing the number of collisions to the number of comets escaping the system yields
the following results:  About 96\% of all comet clones are ejected from the system after
the set integration time of 100 kyr.  Most of these comets escape after a close encounter
with one of the planets.  Taking a closer look on the encounters reveals that 78\% of
the comet clones pass within a tenth of the planet's Hill radius --- but still above the
presumed planetary surface; note that 91\% of those have a close encounter with planet
55~Cnc-d.  The majority of those are ejected from the system as they gain additional
energy as a result of the close planetary encounter.  Moreover, a small fraction of
the comets approach the planet as close as $1/100$ Hill radius (approximately the distance
of the planet's surface); those comets can thus be assumed hitting the planet (collision).
This happens for $\approx$1\% of the total number of comet clones; again, 94\% of the
collisions occur with planet 55~Cnc-d.

%

\begin{table}
	\centering
	\caption{Information on the Remaining Clones of Comet~A}
\begin{tabular}{lccc}
\noalign{\smallskip}
\hline
\noalign{\smallskip}
Number & Semimajor Axis  & Eccentricity & Inclination  \\
...    & (au)            & ...          & ($^{\circ}$) \\
\noalign{\smallskip}
\hline
\noalign{\smallskip}
   1  &   {\p0}2.8035504     &  0.7199375   &      62.0223301 \\
   2  &   {\p0}3.2998323     &  0.6339047   &      32.4305304 \\
   3  &   {\p0}0.9461796     &  0.5210701   &      28.4985011 \\
   4  &       18.4898757     &  0.8552442   &      41.0065245 \\
   5  &   {\p0}7.8064482     &  0.5519021   &      62.1844130 \\
   6  &   {\p0}2.3411522     &  0.5798915   &  {\p0}8.6652167 \\
   7  &   {\p0}0.8607623     &  0.6551015   &      19.0532704 \\
\hline
\end{tabular}
\label{table-Tempel}
\end{table}


\subsubsection{The Comet 9P~/~Tempel~1}\label{examples2}

%

\begin{figure*}
  \includegraphics[width=0.34\linewidth, angle=270]{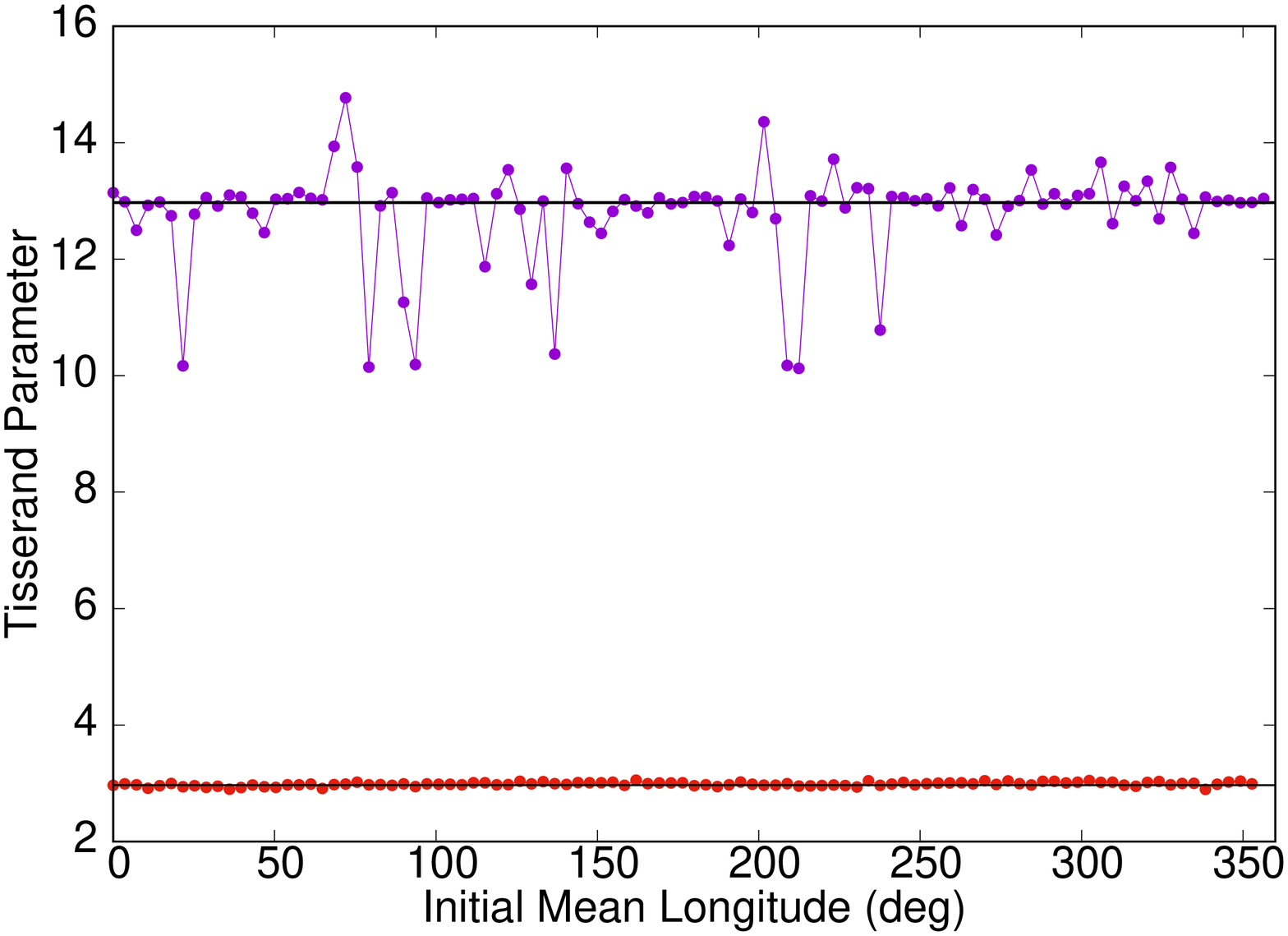} 
  \includegraphics[width=0.34\linewidth, angle=270]{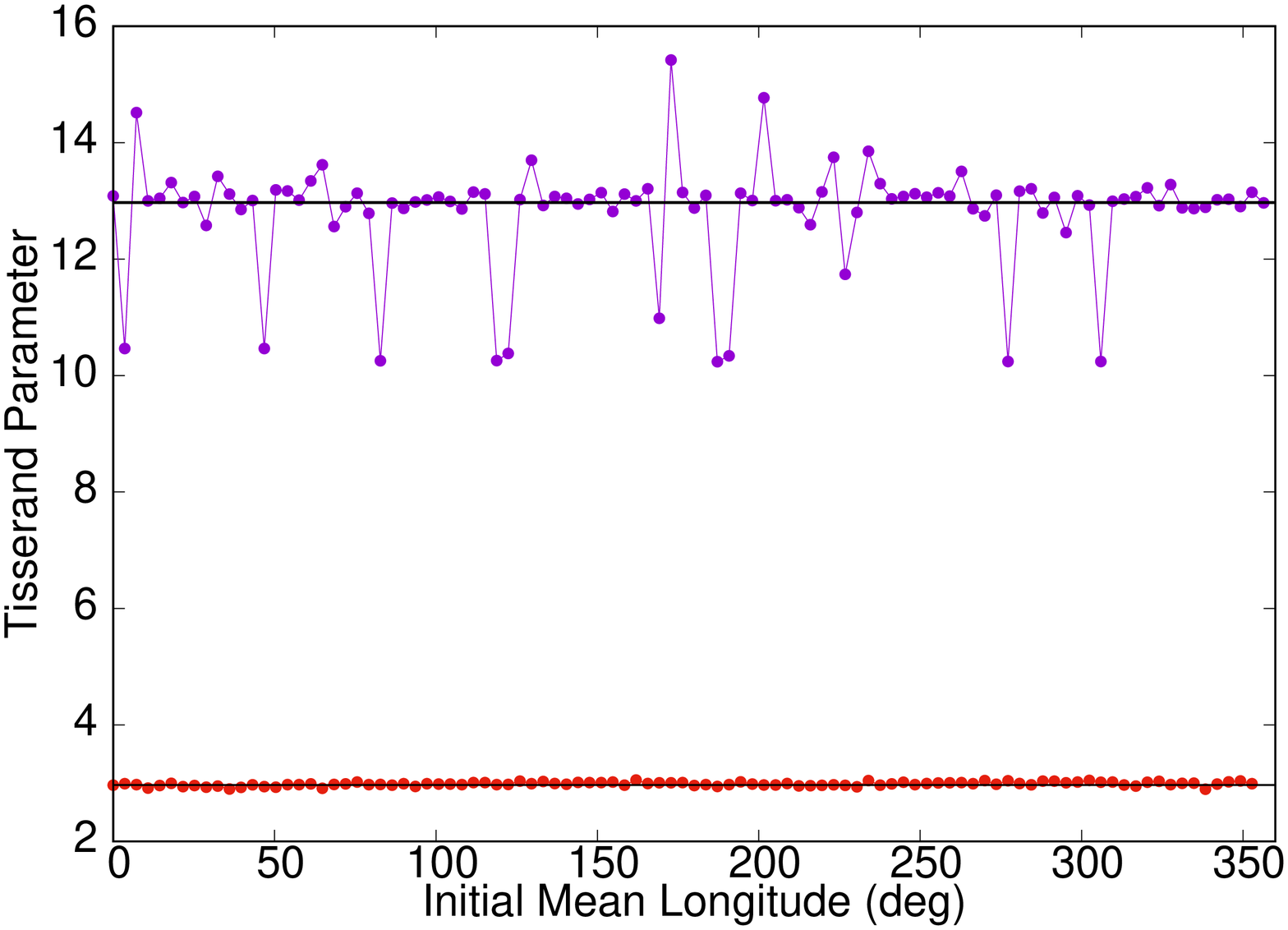}
  \caption{Study of the Tisserand parameter for 9P~/~Tempel~1, represented by 100 clones per model.
    Here we show the results for clones placed into the 55~Cnc system (purple points, values plus 10 for better visibility)
    and the Solar System (red points).  The left panel considers the common system of 55~Cnc (see Table~\ref{tab:planets_param}),
    whereas regarding the right panel a fictitious planet Earth at 1~au has been added.}
  \label{fig:tisserand1}
\end{figure*}

\begin{figure*}
  \includegraphics[width=0.34\linewidth, angle=270]{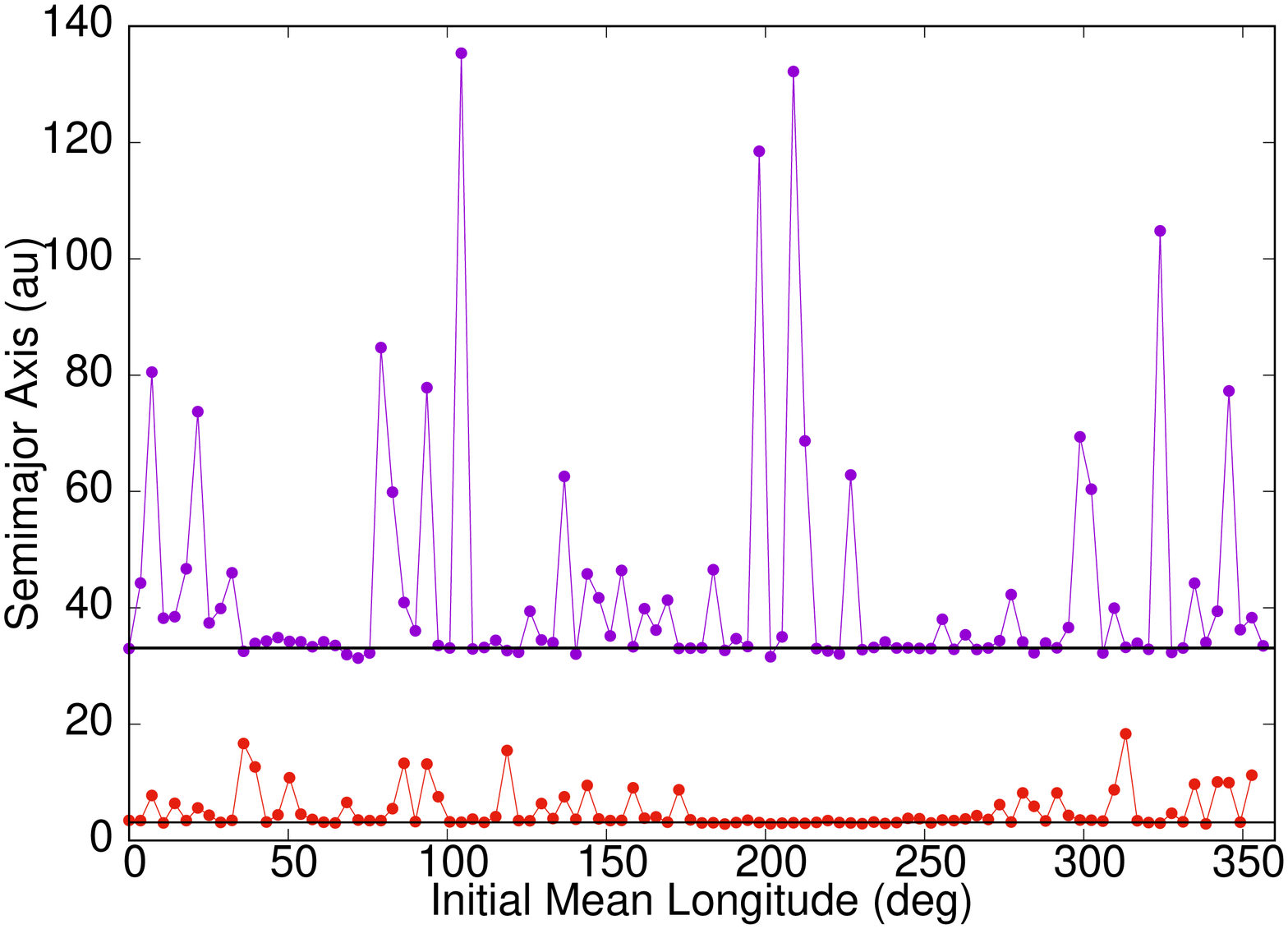} 
  \includegraphics[width=0.34\linewidth, angle=270]{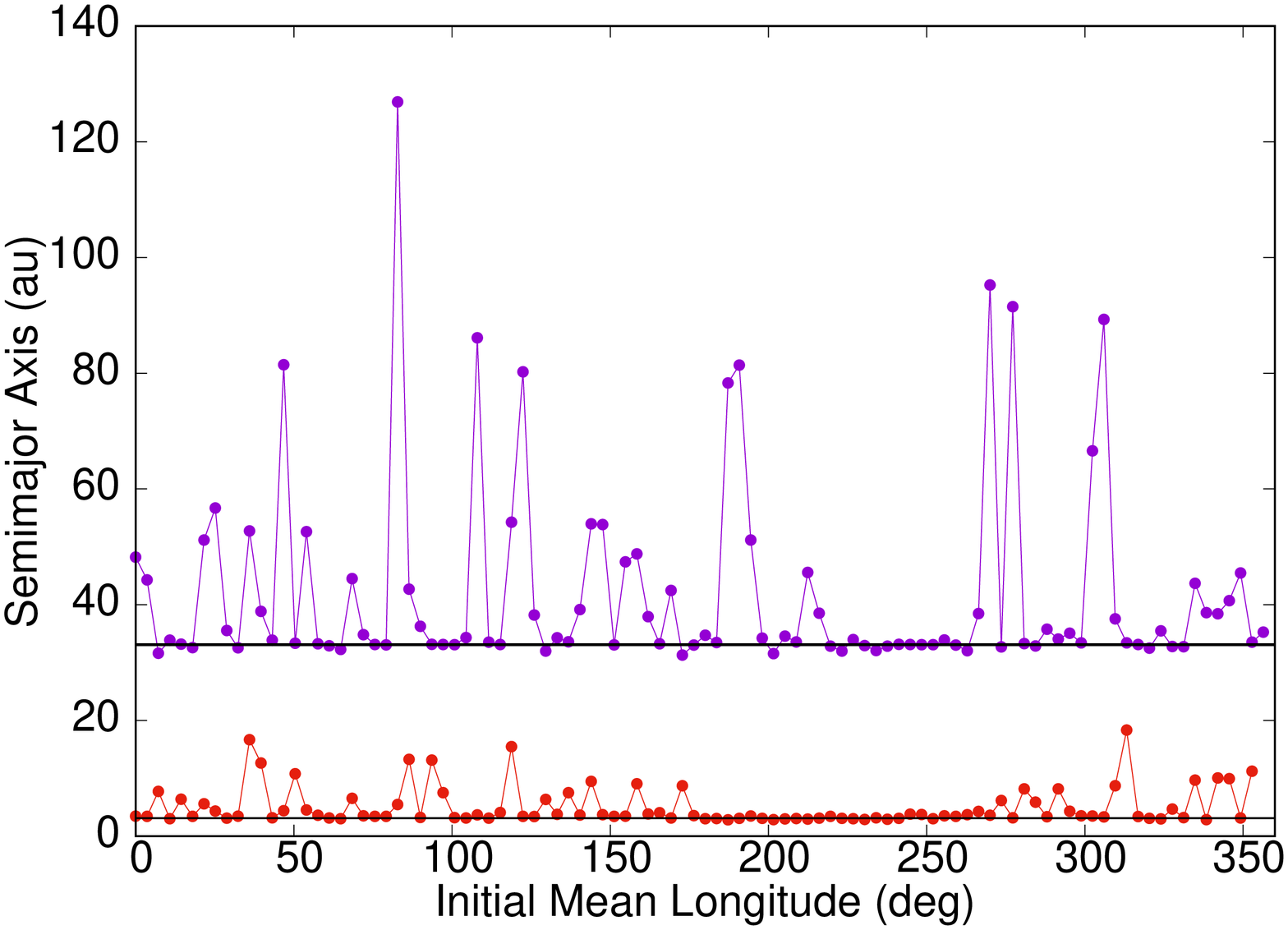}
  \caption{Study of the semimajor axis for 9P~/~Tempel~1, represented by 100 clones per model.
    Here we show the results for clones placed into the 55~Cnc system (purple points, values plus 30 for better visibility)
    and the Solar System (red points).  The left panel considers the common system of 55~Cnc (see Table~\ref{tab:planets_param}),
    whereas regarding the right panel a fictitious planet Earth at 1~au has been added.}
  \label{fig:tisserand2}
\end{figure*}

To further our comparison between the Solar System and 55~Cnc we also focused on the comet 9P~/~Tempel~1, referred
to as comet B, while assuming the following initial conditions: $a=3.14$~au adopted as $a=3.59$~au because of the
larger distance of planet 55~Cnc-d ($a=5.74$~au) compared to Jupiter ($a=5.2$~au).  Furthermore, we assume $e = 0.514$
and $i = 10.5^{\circ}$ (both unaltered).  Subsequently, we compared the results of our simulations of up to 100~kyr
regarding the orbital elements and the Tisserand parameter for the 100 comet clones.  The comparison for the Tisserand
parameters is given in Figure~\ref{fig:tisserand1} and \ref{fig:tisserand2}.  The Earth-mass planet has been included
as before.  Note that the 100 clones of
9P~/~Tempel~1 have been created for the purpose of our numerical simulations through varying the initial
mean anomalies $MA$ while maintaining the same values for all other initial parameters (see Sect. 3.1).

%

Figure~\ref{fig:tisserand1} shows the Tisserand parameter for each comet clone
after an integration time of 100~kyr.  The left panel
depicts the system 55~Cnc using the initial values of table~\ref{tab:planets_param}
without the Earth-mass planet included.  The right panel depicts the same kind of simulation
but with the fictitious Earth-mass planet included.  It is evident that the red dots
--- corresponding to the Tisserand parameter of the 9P~/~Tempel~1 clones in the Solar System --- show
no significant deviations from the value T=3.  However, in the system of 55~Cnc some of the
comet clones experience orbital changes resulting in significantly different numbers.  For better
visibility we increased the Tisserand parameters for each clone (purple points) by a factor of 10.

Figure~\ref{fig:tisserand2} depicts the semimajor axes of the various comet clones.  Again, the
left panel depicts the results without the fictitious Earth-mass planet.  Here some
comet clones within 55~Cnc experience significant orbit changes pushing them out
to orbits with high values for the semimajor axis.  In contrast, regarding the Solar System comet
clones with initial mean anomalies between 170$^{\circ}$ and 270$^{\circ}$ remain within their
initial orbits while maintaining low values for semimajor axis. 

\begin{figure*}
  \includegraphics[width=0.43\linewidth, angle=270]{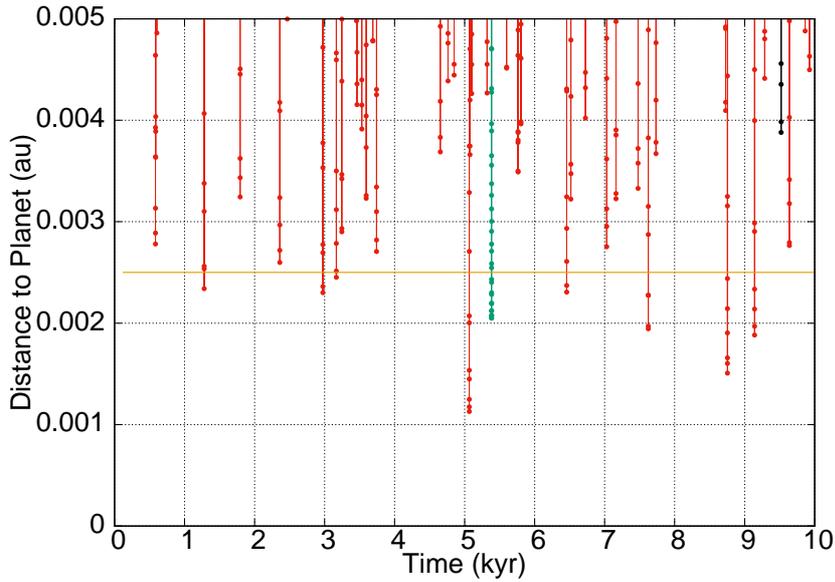} 
  \caption{Close encounters of the clones of 9P~/~Tempel~1 with the Earth in our Solar System.
    The $y$-axis shows the distance of the object from the gravitational centre of the body undergoing a close encounter.
    The yellow line at 0.0025 denotes the distance of the Moon from the gravitational centre of the Earth.
    The red lines depict the distances for a clone from Mars, whereas the green line shows a close encounter
    of a clone with Jupiter. The black line shows the approach of a clone towards Earth.}
  \label{fig:Tempelclose}
\end{figure*}

\begin{figure*}
  \includegraphics[width=0.34\linewidth, angle=270]{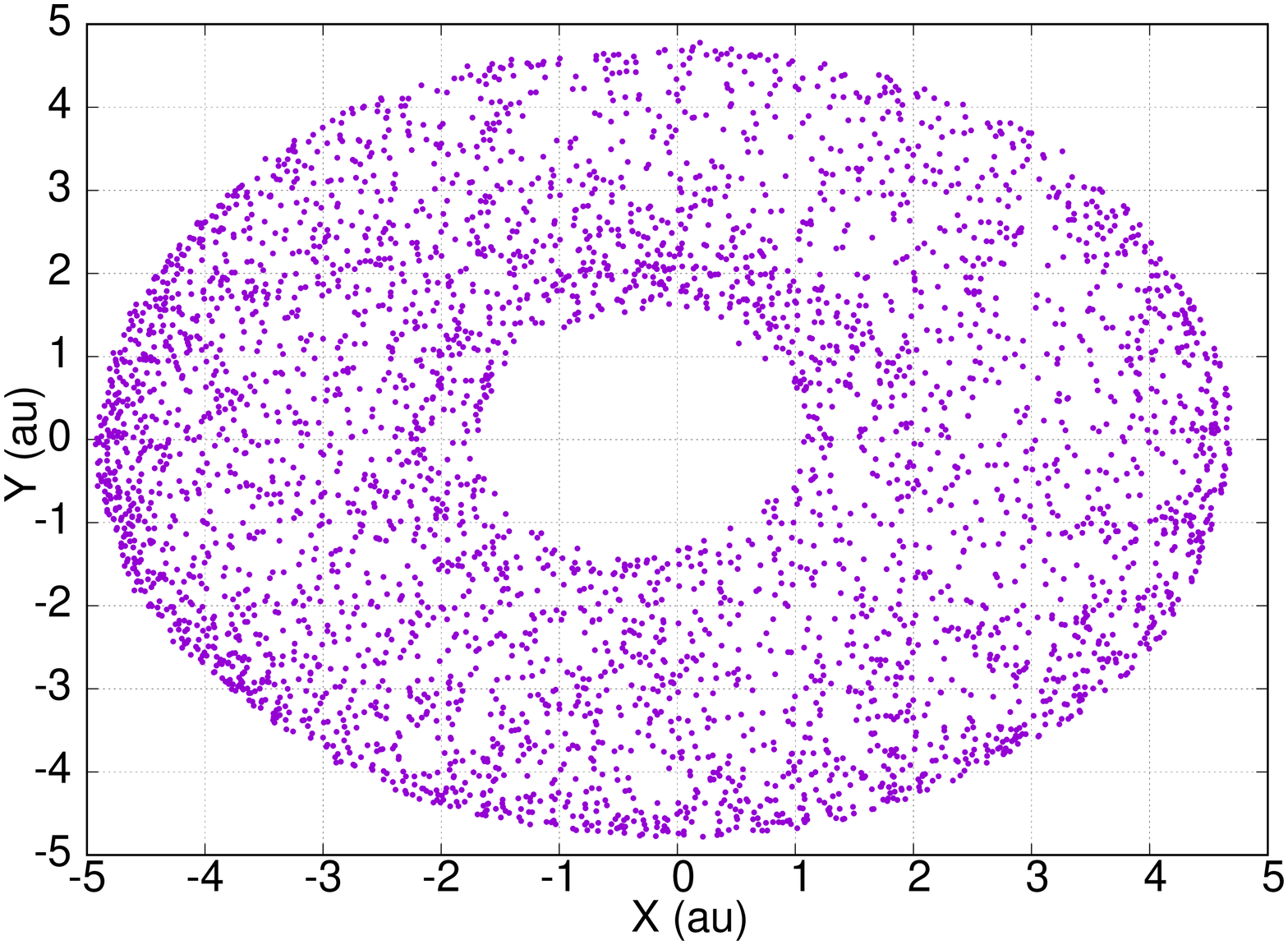}
  \includegraphics[width=0.34\linewidth, angle=270]{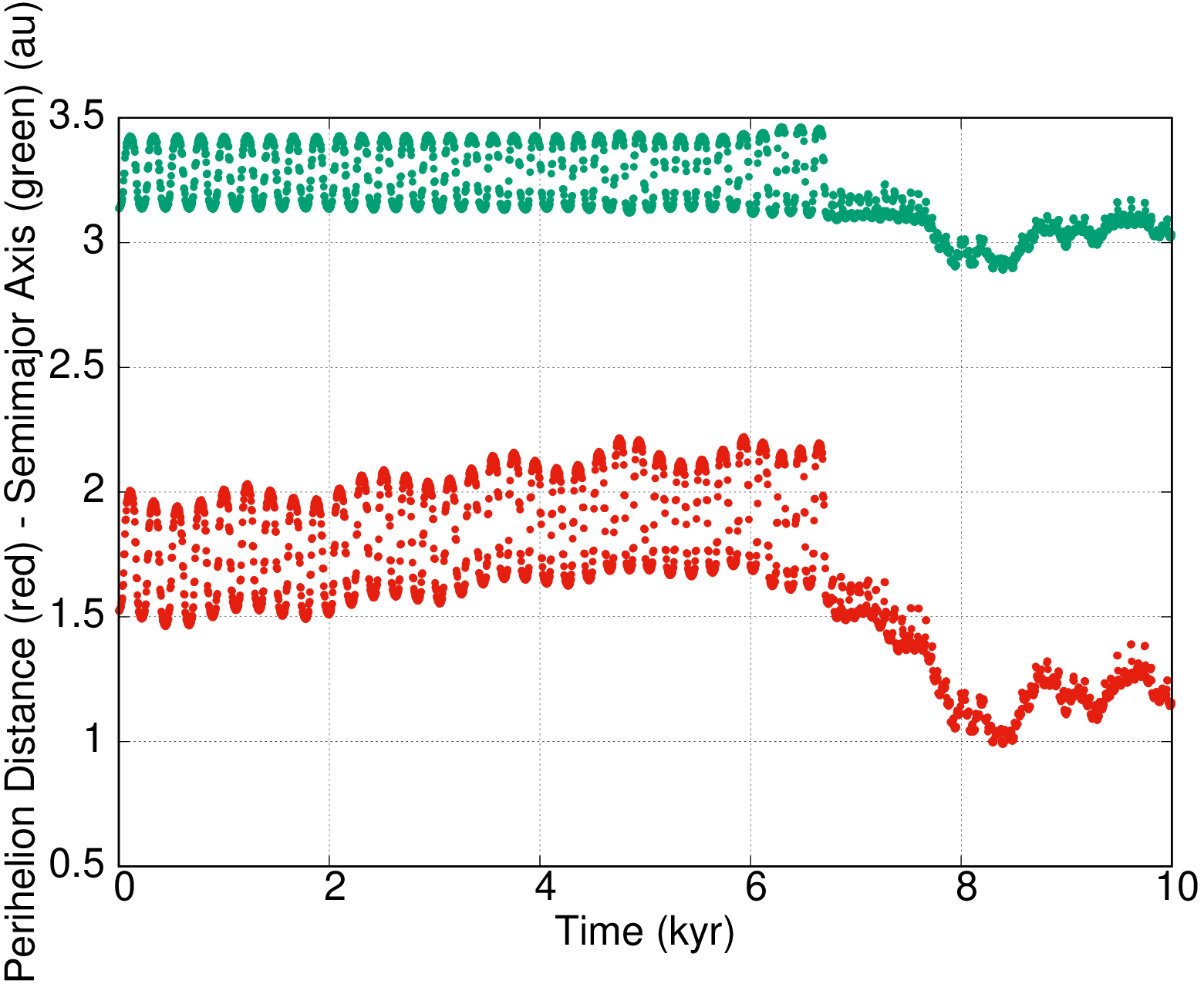}
  \caption{Left: Cometary orbits regarding 9P~/~Tempel~1 in our Solar System with the $x$ and $y$-axis in au
    and the Sun situated at the centre at (0,0).  Right: The red line corresponds to the perihelion distance
    of the comet clone, whereas the green line corresponds to its semimajor axis.  It is found that it crosses
    Mars's orbit (1.5 au) and at about 8 kyr it even reaches Earth's orbit (1 au).}
  \label{fig:Tempelorbit1}
\end{figure*}

We also re-investigated the orbit of 9P~/~Tempel~1 in the Solar System and found that
some comet clones experience close encounters with the Earth, see Figure~\ref{fig:Tempelclose}.
The yellow line corresponds to the distance of the Moon from the gravitational centre of the Earth (0). 
Figure~\ref{fig:Tempelorbit1} shows the orbit in the in the $x-y$ plane of 9P~/~Tempel~1 in our Solar System.
Evidently, close encounters with Earth or Mars are expected to happen during the comet's orbital evolution, but no encounters
with Venus are possible.

The time evolution of the perihelion and semimajor axis for 9P~/~Tempel~1 are shown in Figure~\ref{fig:Tempelorbit1}.
This comet starts at an orbit with a perihelion distance close to Mars's orbit (red line in Figure~\ref{fig:Tempelorbit1}).
After about 7~kyr, changes in the semimajor axis occur leading to a decrease in the pericentre
distance from the Sun.  The comet crosses Earth's orbit at 1~au after about 8~kyr.  However,
a close approach to Earth happens only after about 9.5~kyr.

%

%

\section{Conclusions and Further Research}\label{sec:conclusions}

The aim of this study was to explore the behaviour of fictitious JFCs in the
55~Cnc star--planet system.  55~Cancri (or, technically, 55~Cnc~A) bears some resemblance to the Sun,
although various differences exist.  It has a mass of 0.96~$M_\odot$ \citep{lig16}, which
is lower than the solar mass.  Furthermore, its effective temperature and luminosity are below
the solar values as well.  Moreover, 55~Cnc is known to host five planets.  All of them
have masses significantly larger than Earth.  In addition, following observational constraints,
there is a gap without planets between $\sim$0.8~au and $\sim$5.7~au, offering the principle
possibility of habitable terrestrial planets, including the prospect of long-term orbital stability.

What is the main result of our investigation pertaining to the orbital integration of $\sim$60,000
fictitious JFCs in 55~Cnc implemented through the study of comet clones?
In order to answer this question, each comet clone has been integrated for 100~kyr.  Comet clones
have been considered for 2P~/~Encke, 9P~/~Tempel~1, and 31P~/~Schwassmann-Wachmann~2.

In conclusion, we found that the duration of stability for JFC analogs in 55~Cnc is much
less compared to the corresponding comets of the Solar System.  This outcome occurs because of the profound differences
in the structure of the planetary system of 55~Cnc relative to the Solar System.  Between 55~Cnc-f
and 55~Cnc-d, located at 0.781~au and 5.74~au, respectively, the comets in 55~Cnc do not undergo
close planetary encounters akin to Solar System comets with respect to Earth and Mars.  The reason
is that the planets of the 55~Cnc system possess a significantly higher mass compared to the
Solar System planets; consequently, the comets in the 55~Cnc system exhibit a lower level of
orbital stability.  On the other hand, an increased number of encounters and collisions would
be expected between comets and terrestrial planets regarding the 0.8 / 5.7~au gap.

The region between 0.8~au and 5.7~au allows terrestrial planets to exhibit long-term orbital
stability and potential habitability; in fact, this region also encompasses 55~Cnc's habitable zone
\citep[e.g.,][]{ray08,sat19}.  Therefore, comets entering this domain are expected to be relevant for
the proliferation of water and prebiotic matter to be provided through comet--planet collisions.
Hence, there is a need for targeted future observations, especially through the usage of large telescopes,
as, e.g., the James Webb Space Telescope \footnote{The James Webb Space Telescope
(JWST) is a large, space-based observatory optimized for infrared wavelengths for observing terrestrial 
planets but with many also many other scientific goals.  It is scheduled to be launched in 2021.}.
The overall structure of 55~Cnc's planetary system, including the possible impact of comets,
shall be considered a prodigious inspiration for future theoretical work.


\section*{Acknowledgements}

This research is supported by the Austrian Science Fund (FWF) through grant S11603-N16 (R.~D. and B.~L.).
Moreover, M.~C. acknowledges support by the University of Texas at Arlington.  The computational results
presented have been achieved in part using the Vienna Scientific Cluster (VSC).





%
%


\bsp	
\label{lastpage}
\end{document}